\documentclass[twocolumn,twoside,slac_two]{revtex4-1}
\usepackage{graphicx}
\usepackage{fancyhdr}
\usepackage{hyperref}
\hypersetup{
    colorlinks=true,
    urlcolor=magenta,
}
\begin{document}

\title{Influence of the proton initiated at most two electromagnetic sub-cascades events on IACT observations}

%

\author{Dorota Sobczynska, Katarzyna Adamczyk}
\affiliation{University of {\L }\'{o}d\'{z}, 
Department of Astrophysics, Pomorska 149/153, 90-236 {\L }\'{o}d\'{z}, Poland}

\begin{abstract}
The efficiency of the $\gamma$/hadron separation worsens significantly at low energies for Imaging Air Cherenkov Telescopes (IACT). This observed effect was partially explained by the occurrence of a hardly reducible hadronic background (i.e. detected images that are formed mainly by the light from a single electromagnetic or a single $\pi^0$ sub-cascade in the proton induced shower) \cite{sob2009b,sob2015}. IACTs also record events containing Cherenkov light from at most two electromagnetic sub-cascades, which don't have to be products of single $\pi^0$ decay in the hadron initiated showers. In this paper, the impact of at most two electromagnetic sub-cascades events on the primary $\gamma$-ray selection was studied using the Monte Carlo simulations. We investigate how the total fraction of at most two sub-cascades events in the expected total protonic background depends on the hadron interaction models (GHEISHA and FLUKA), trigger threshold, reflector area and altitude of the observatory. We show that the efficiency of the $\gamma$/hadron separation is anti-correlated with the contribution of at most two electromagnetic sub-cascades events in the proton initiated showers below 200 GeV. The influence of at most two electromagnetic sub-cascades events on the $\gamma$-ray selection is similar to the effect of the single electromagnetic sub-cascade and singe $\pi^0$ events in IACT observations. The occurrence of all these images in the data collected by IACTs reduces the efficiency of the $\gamma$/hadron separation at low energies.
\end{abstract}

\maketitle

\thispagestyle{fancy}


\section{Introduction}

The Air Imaging Cherenkov Telescopes are used to detect the primary $\gamma$ rays from the cosmic sources. The sensitivities of IACT systems becomes worse below 100 GeV (see e.g. \cite{perf14b}). The decrease of the $\gamma$/hadron separation efficiency at low energies may be explained by several reasons: {\bf a)} due to the larger fluctuations of the Cherenkov light density at ground \cite{sob2009} one may expect larger fluctuations of the image parameters; {\bf b)} the geomagnetic field influences on the image parameters \cite{comm08,szan13}; {\bf c)} the existence of the images of the primary electron or positron \cite{elhess}; {\bf d)} IACTs may be triggered by Cherenkov photons produced by charged particles from a single or at most two electromagnetic sub-cascades (which are a products of a single $\pi^{0}$ decay) in the hadron initiated shower \cite{sob2007,sob2009b,sob2015}.

In this paper we investigated the effect which is similar to the single $\pi^{0}$ events. We consider registered images that contain most of Cherenkov light from at most two electromagnetic (em) sub-cascades, which don't have to be a products of single particle decay in the proton induced shower. Such events are hardly reducible background in IACT observations because they have similar shapes to primary $\gamma$-ray images. On the one hand at most two em sub-cascades images may be slightly wider as the primary $\gamma$-ray events because a separation angle between two secondary $\gamma$'s is expected. On the other hand, both em sub-cascades start deeper in the atmosphere and by that each of them has the narrower distribution of charged particles in comparison to the shower initiated by primary $\gamma$ rays. As the result the primary $\gamma$-rays and two em sub-cascades images are similar and a deterioration of the $\gamma$/hadron separation efficiency is expected. 

All results presented in this paper are based on Monte Carlo (MC) simulations. We demonstrate how the total fraction of two em sub-cascades events in the expected protonic background depends on the hadron interaction model (GHEISHA and FLUKA) and chosen IACT features like the trigger threshold, multiplicity of triggered IACTs, telescope size and altitude of the observatory. We show that the contribution of at most two em images decreases with the energy of primary proton and with the average SIZE of detected images. 
Two Hillas parameter that describe the image shape (the scaled WIDTH and scaled LENGTH \cite{daum97}) are used in the $\gamma$/hadron separation. We present the strong anti-correlation between the calculated quality factor and the contribution of two em sub-cascade events (before a primary $\gamma$-rays selection). This fact indicates that two em sub-cascades events also cause a deterioration of the $\gamma$/hadron separation efficiency at low energies.

\section{Monte Carlo Simulations}
Two versions of the CORSIKA code \cite{heck} were used for MC simulations of the Extensive Air Shower (EAS) development in the atmosphere. In the 6.023 version GHEISHA \cite{gheisha} and VENUS \cite{venus} interaction models have been applied for the low (i.e. for particles with primary momentum below 80~GeV/c) and high energy ranges. The FLUKA \cite{fluka} and QGSJET-II {\cite{ostap06a}} have been used as the low and high energy interaction models in the version 6.99. We have modified the standard code in order to have additional information about each em sub-cascade created in the shower.

Four 230~$m^{2}$ telescopes (MAGIC-like) placed in the corners of a diamond were chosen as an example of IACT system. The side length of the diamond was fixed to 85~m and diagonals to 85~m and 147~m for the simulated altitude of the observatory equal to 2200~m a.s.l.. All distances between IACTs were reduced by a factor of 0.75 for the altitude of 4~km a.s.l.. 
The night sky background and the geomagnetic field were fixed to the MAGIC site (La Palma). Five sets of MC have been performed. The input parameters which were used for all MC simulations are presented in Table~1. The differences between simulation sets are shown in Table~2. The detailed information about the simulated system can be found in \cite{sob2015}.

\begin{table}[t]
\begin{center}
\caption{Input parameters of the MC simulations.}
\begin{tabular}{|l|c|c|}
\hline \textbf{} & \textbf{Proton} & \textbf {$\gamma$ ray} 
\\
\hline E$_{min}$ [GeV] & 30 & 10 \\
\hline E$_{max}$ [TeV]& 1 & 1 \\
\hline Spectral index & -2.75 & -2.6 \\
\hline Zenith angle [deg.] & 20 & 20 \\
\hline Azimuth angle [deg.] & 0 & 0\\
\hline Full opening angle of cone [deg.] & 11 & 0 \\
\hline Max. of impact parameter [m] & 1200 & 350 \\
\hline number of events [10$^{6}$] & 22 & 1 \\
\hline
\end{tabular}
\label{input_par}
\end{center}
\end{table}

\begin{table}[t]
\begin{center}
\caption{The differences between simulation sets. Interaction models G/V and F/Q correspond to GHEISHA/VENUS and FLUKA/QGSJET-II, respectively.}
\begin{tabular}{|l|c|c|c|c|c|}
\hline \textbf{MC set} & \textbf{ I } & \textbf{ II } & \textbf{ III } & \textbf{ IV } & \textbf{ V}\\
\hline \textbf{Interaction} & \ { G/V } & { G/V } & { G/V } & { G/V } & { F/Q }\\
\textbf{Models} & & & & &\\
\hline \textbf{Refl. area [m$^{2}$]} & ~~230~~& ~~230~~ & ~~160~~ & ~~100~~ & ~~230~~\\
\hline \textbf{altitude [km]} & 2.2 & 4.0 & 2.2 & 2.2 & 2.2 \\
\hline
\end{tabular}
\label{l2ea4-t1}
\end{center}
\end{table}

\section{Results and Discussions}
Two em sub-cascades events (or images), which are investigated in this paper, are defined as triggered showers that have more than 90$\%$ of total SIZE formed by Cherenkov light from at most two the same em sub-cascades in each recorded image. The rest may be by muonic and hadronic part of the shower. 

The ratio between the number of two em sub-cascades events and the number of all triggered proton showers versus the primary energy is presented in Figure~1. This ratio was calculated separately in each histogram bin for the trigger threshold of 3~photoelectrons (phe.) and the condition that all recorded SIZEs larger than 10~phe. Two possible required trigger conditions: at least one and at least two triggered IACT are plotted in panels a and b, respectively. The contribution of the hardly reducible background in the proton initiated showers decreases with the energy. The presented fraction is the highest for the IACT system located at higher altitude and the lowest for the smallest area of the telescope mirror. We have checked that more then 90$\%$ of two em sub-cascades events are proton initiated showers with primary energy below 200 GeV, regardless on the interaction model, altitude of the observatory, telescope size and trigger conditions.

\begin{figure}
\includegraphics[width=80mm]{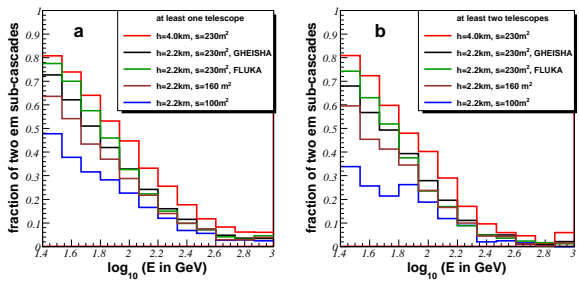}
\caption{The ratio of the number of two em sub-cascade events to the number of all triggered proton shower versus primary energy for the trigger threshold equal to 3~phe. and with the SIZE parameter of each recorded image larger than 10~phe.: {\bf a)} at least one; {\bf b)} at least two triggered IACTs.}
\label{energy}
\end{figure}

In order to estimate the expected total fraction of two em sub-cascades events in the total proton background, a simple power-law fits of the energy distribution tail has been used to evaluate the expected number of the triggered events with energies above 1 TeV. The fraction of two em sub-cascades events in the proton background is shown in Figure~2. The results obtained from the same MC set show that, the presented fraction decreases with the trigger threshold and with the multiplicity of the triggered telescopes (from one to three in panels a, b and c, respectively). Both higher trigger threshold and higher multiplicity correspond to higher primary energies that less probably appear to be two em sub-cascades images. Because of the same reason higher fractions of hardly reducible background are expected for the system located at higher altitudes and for larger telescope sizes.
FLUKA/QGSJET-II model gives higher fraction of two em sub-cascades images than GHEISHA/VENUS for at least one triggered telescope only. Quantitatively, the difference is the same as in \cite{sob2015} where only the single em sub-cascade images were considered.

\begin{figure}
\includegraphics[width=80mm]{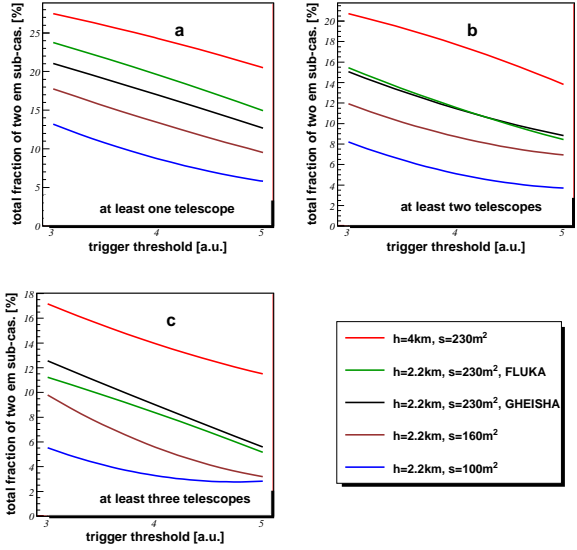}
\caption{ Expected fraction of two em sub-cascades events in the protonic background as a function of the trigger threshold for: {\bf a)}~at least one; {\bf b)}~at least two; {\bf c)}~at least three triggered telescopes.}
\label{fraction}
\end{figure}

 Figure~3 show how the ratio of the number of two em sub-cascades events to the number of all triggered protons decreases with the average SIZE for the trigger threshold of 3~phe. and SIZE parameter of all recorded images larger than 10~phe.. The higher fraction is expected for the simulated altitude of 4~km than for 2.2~km in the fixed $<$SIZE$>$ bin. The contribution of the hardly reducible background increases with the telescope area as expected. The differences between both investigated interaction models are negligible in this Figure. More than 95$\%$ of two em sub-cascades images have the $<$SIZE$>$ lower than 300~phe., regardless on the interaction model, altitude of observatory, telescope size and trigger conditions. 

\begin{figure}
\includegraphics[width=80mm]{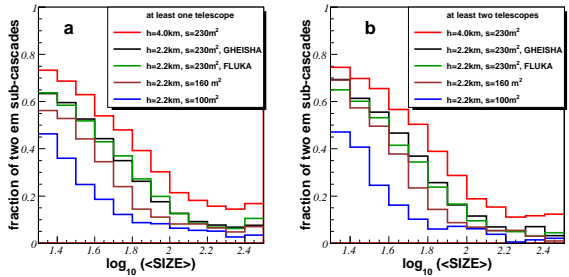}
\caption{The ratio of the number of two em sub-cascades events to the number of all triggered proton shower versus $<$SIZE$>$ for the trigger threshold equal to 3~phe.: {\bf a)}~at least one; {\bf b)}~at least two triggered telescopes. The ratio was calculated in each histogram bin separately.}
\label{size}
\end{figure}
 
The so-called scaled WIDTH (WIDTH$_{S}$) and scaled LENGTH (LENGTH$_{S}$) \cite{daum97} have been used in the method of the $\gamma$/hadron separation. Both of those parameters describe the image shape.
Figures~4a and 4b show the distributions of average WIDTH$_{S}$ and LENGTH$_{S}$ for at least two triggered telescopes at the trigger threshold equal to 3~phe.. The histograms of all triggered events for primary $\gamma$ ray (highest) and proton (dashed) are normalised to~1. Solid histograms present two em sub-cascades images. 
It is seen in Figure~4, that the primary $\gamma$ rays and investigated background have similar distributions. That is why two em sub-cascades events are hardly reducible background in the $\gamma$/hadron separation.

\begin{figure} [t]
\begin{center}
      \includegraphics[width=40mm]{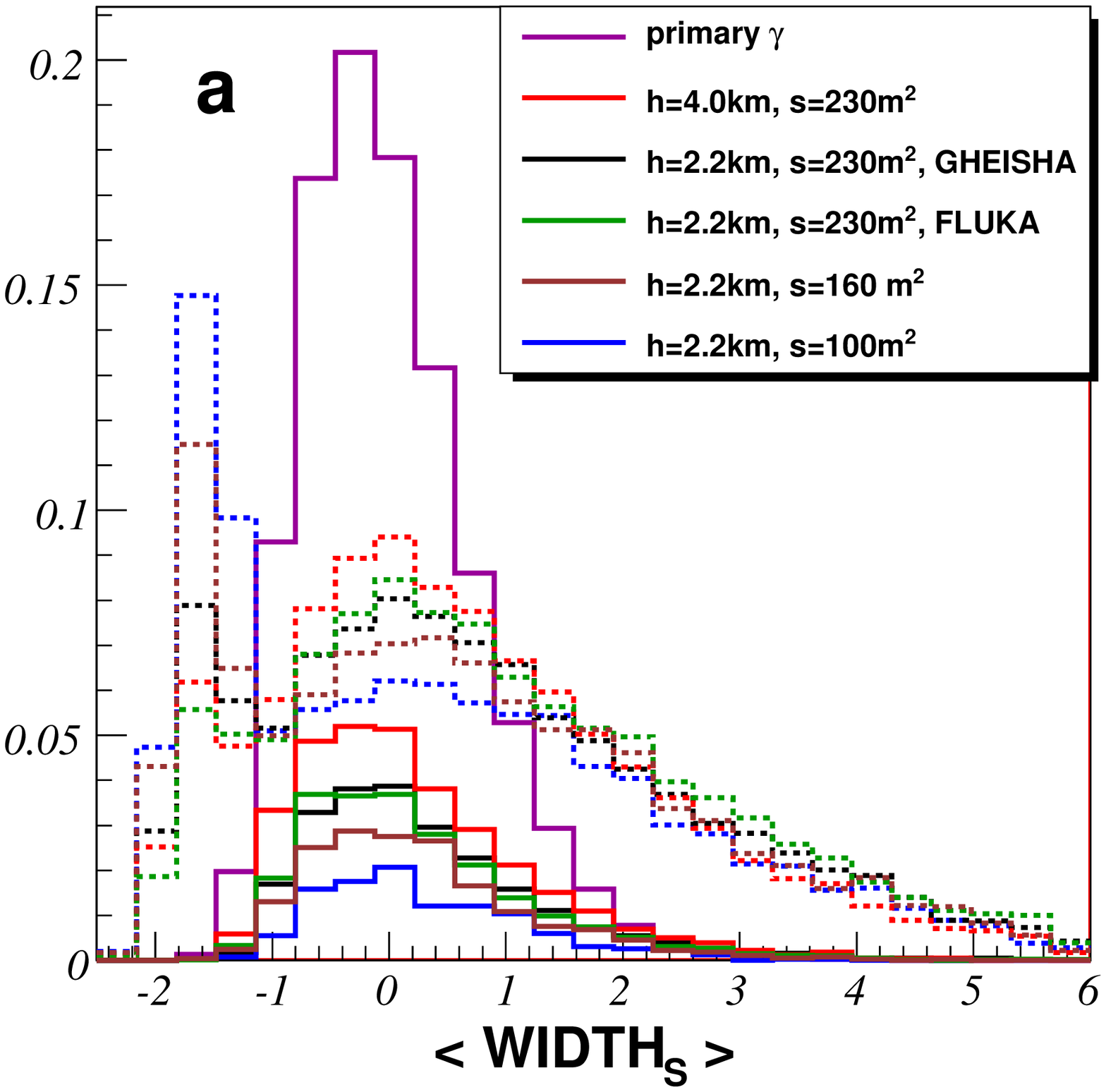}
      \includegraphics[width=40mm]{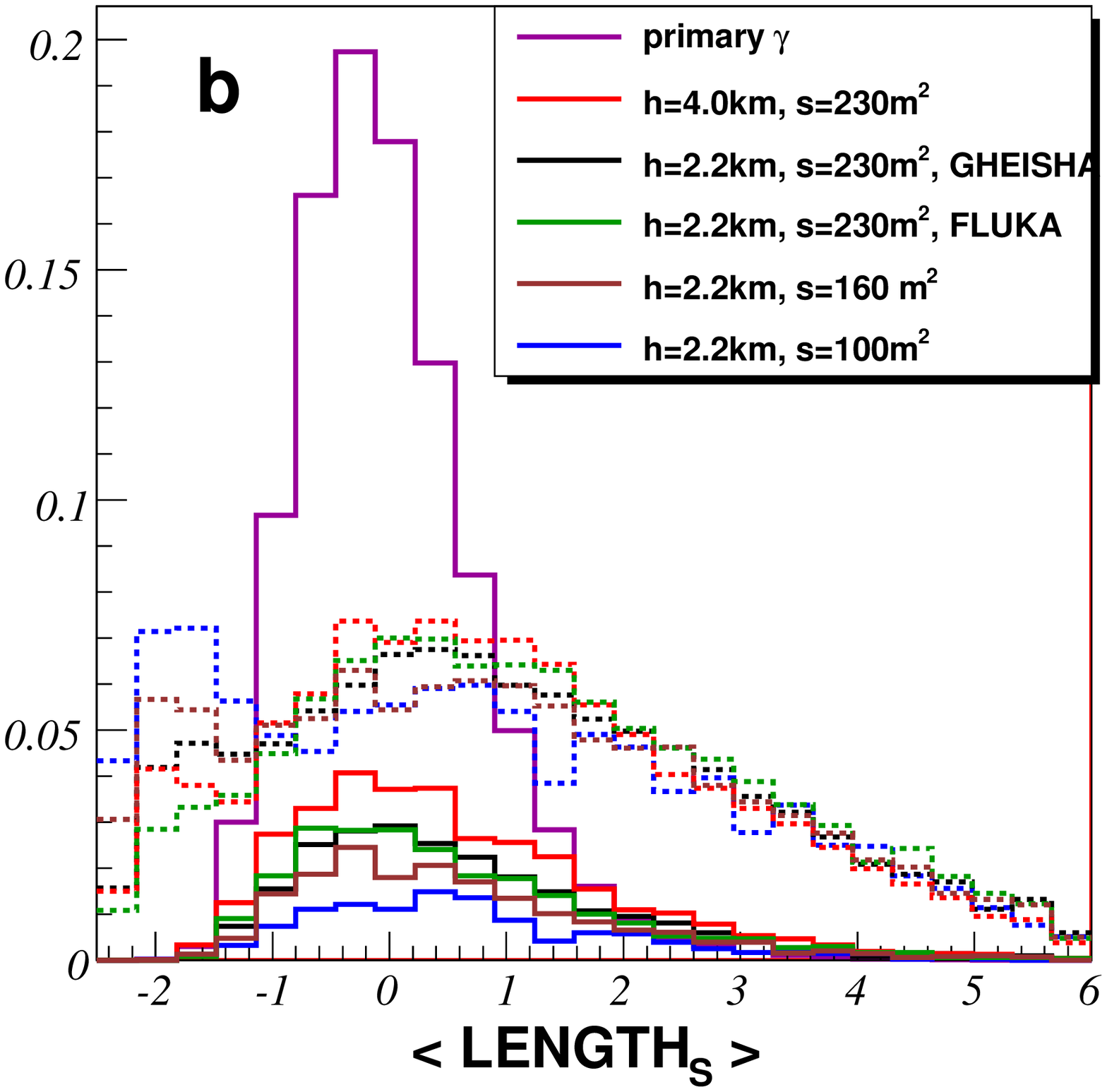}

\caption{{\bf a)} The mean WIDTH$_{S}$ distribution. {\bf b)} The mean LENGTH$_{S}$ distribution. Dashed and solid histograms correspond to all triggered and two em sub-cascades events for primary protons, respectively.}
\label{wid_len}
\end{center}
\end{figure}

The $\gamma$/hadron separation method was applied to the MC data in five intervals of $<$SIZE$>$. 
Events that have all recorded images with WIDTH$_{S}$ and LENGTH$_{S}$ in the range between -1.5 and 1.5 were chosen as primary $\gamma$-ray candidates. In order to demonstrate the efficiency of the $\gamma$-ray selection from the hadronic background we used the quality factor (QF) \cite{aha93}.
 The contribution of two em sub-cascades events in the total protonic background was calculated before the primary $\gamma$-ray selection. The dependence between QF and the contribution of investigated events is shown in Figure~5 for at least two triggered IACTs at the trigger threshold of 4~phe.. The quality factor decreases with the fraction of two em sub-cascades events for all simulation sets and all investigated trigger conditions. 

\begin{figure}
\includegraphics[width=60mm]{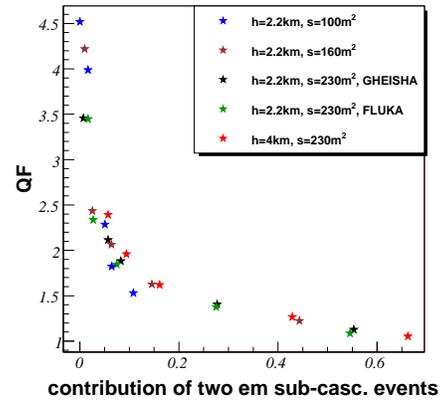}
\caption{The quality factor versus the contribution of two em sub-cascades events in proton background before the $\gamma$/hadron separation for at least two triggered IACTs at the trigger threshold of 4~phe.}
\label{quality_factor}
\end{figure}

\begin{table}[t]
\begin{center}
\centering
\caption {\label{tab1} Correlation coefficients between the QF and the contribution of two em sub-cascades events before the $\gamma$-ray selection for a trigger threshold of 3~phe..}
\begin{tabular}{|l|c|c|c|c|c|}
\hline \textbf{MC set} & \ { I } & { II } & { III } & { IV } & { V}\\
\hline
1 trigg. IACT &-0.98&-0.99&-0.97&-0.88&-0.96\\
2 trigg. IACTs&-0.81&-0.93&-0.75&-0.76&-0.83\\
3 trigg. IACTs&-0.70&-0.88&-0.60& &-0.82\\
4 trigg. IACTs&-0.60&-0.85& & &-0.76\\
\hline
\end{tabular}
\label{l2ea4-t1}
\end{center}
\end{table}

The calculated correlation coefficients (based on five points) between the quality factor and the contribution of two em sub-cascades events are presented in Table 3 for the trigger threshold of 3~phe.. Those coefficients are slightly worse for higher trigger thresholds. The coefficients were not calculated if in at least one $<$SIZE$>$ bin any of protonic showers survived $\gamma$-ray selection. The strong anti-correlation between the $\gamma$/hadron separation efficiency and the contribution of two em sub-cascades events indicates that this specific kind of proton induced showers is one of the main reasons of the worsening of the primary $\gamma$-ray selection efficiency at low energies.

\section{Conclusions}
Our results show that proton images containing majority of the light from at most two em sub-cascades can be detected by systems of IACTs with the size of a single telescope between 100~$m^{2}$ and 230~$m^{2}$. The vast majority of this background has a small primary energy (below 200~GeV) and a small detected $<$SIZE$>$ (below 300~phe.), regardless on the interaction models, altitude of the observatory, mirror area, and investigated trigger conditions.

The estimated fraction of two em sub-cascades events in the protonic background decreases with the trigger threshold and the number of triggered telescopes. This fraction is lower for the altitude of 2.2~km than for 4.0km a.s.l., (for fixed trigger conditions). The fraction of hardly reducible background increases with the mirror area of the single telescope. 
FLUKA/QGSJET-II model gives higher fraction of two em sub-cascades images than GHEISHA/VENUS for at least one triggered telescope only. Quantitatively, the difference is the same as in \cite{sob2015} where only the false $\gamma$-ray events (single em sub-cascade images) were considered.

The contribution of two em sub-cascades events in protonic background that is calculated in the $<$SIZE$>$ interval diminishes with the average SIZE, for all simulations sets. This contribution increases with the altitude of the observatory and telescope size for fixed trigger conditions. Both tested interaction models gives similar contribution of two em sub-cascades events in in the proton initiated showers.
 
Images of two em sub-cascades and primary $\gamma$ ray have similar shapes. Therefore, the $\gamma$-ray selection method based on WIDTH and LENGTH parameters is not effective if the contribution of two em sub-cascades events in protonic background is high. We have found and shown the strong anti-correlation between the efficiency of the $\gamma$/hadron separation and the contribution of two em sub-cascades events (before $\gamma$-ray selection). Additionally, for the IACT system triggering by higher energy protons (e.g smaller reflector size and/or higher multiplicity of triggered IACTs) the this correlation becomes weaker.

 The occurrence of two em sub-cascades events is also the reason of the degradation of both the efficiency of the $\gamma$/hadron separation and the sensitivity of the IACT system at low energies. 

\begin{acknowledgments}
D. Sobczy\'{n}ska was supported by Polish MNiSzW grant 745/n HESS MAGIC 2010/0. Katarzyna Adamczyk was supported by Polish MNiSzW grant for young scientists number B1611500001134.02.
\end{acknowledgments}


\end{document}